\documentstyle[PASJadd]{PASJ95}
\begin{document}

\setcounter{page}{1}

\title{Near-Infrared Spectroscopy of the Cool
Brown Dwarf, SDSS 1624+00
}

\author{Tadashi {\sc Nakajima} \\
{\it National Astronomical Observatory, 2-21-1 Osawa,
Mitaka, 181-8588} \\
{\it E-mail(TN): tadashi.nakajima@nao.ac.jp}
\\[6pt]
Takashi {\sc Tsuji} \\
{\it Institute of Astronomy, University of Tokyo,
2-21-1 Osawa, Mitaka, 181-8588} \\[6pt]
Toshinori {\sc Maihara}, Fumihide {\sc Iwamuro}, Ken-taro {\sc Motohara},
Tomoyuki {\sc Taguchi}, 
Ryuji {\sc Hata} \\
{\it Department of Physics, Kyoto University, Kitashirakawa, 
Kyoto 606-8502} \\[6pt]
Motohide {\sc Tamura} \\
{\it National Astronomical Observatory, 2-21-1 Osawa,
Mitaka, 181-8588} \\
and \\
Takuya {\sc Yamashita} \\
{\it Subaru Telescope, National Astronomical Observatory
of Japan, 650 North A$^\prime$ohoku Place Hilo,
Hawaii 96720, USA}
}

\abst{
Using the Subaru Telescope,
we have obtained multiple near-infrared spectra
of the cool brown dwarf,  SDSS 1624+00 (J162414.37+002915.8),
in search of spectral variability in an 80 minute time span.   
We have found the suspected variability of water vapor absorption
throughout the observations, which requires confirmation
with a longer time baseline.
After coadding the spectra, we have obtained
a high-quality spectrum covering from 1.05 to 1.8 $\mu$m.
There are three kinds of spectral
indicators, 
the water vapor bands, methane band and  K I lines
at 1.243 and 1.252 $\mu$m, which can be used
to study temperature and the presence of dust. 
We compare 
the spectra of  SDSS 1624+00 and Gliese 229B, paying
special attention to these indicators. 
The shallower water vapor absorption of SDSS 1624+00
indicates that it is warmer and/or dustier. 
The shallower methane absorption suggests that
SDSS 1624+00 is warmer. 
We interpret
the deeper K I lines  in
SDSS 1624+00  as the result of its higher temperature.
With the help of model spectra, we conclude that
SDSS 1624+00 is warmer and dustier than Gliese 229B.
For the first time in a cool brown dwarf,  a finite flux
is seen at the bottom of the water vapor band between 1.34 and
1.42 $\mu$m, which means that the 1.4 $\mu$m band of water
can completely be observed from the ground.
}

\kword{Spectroscopy --- Stars: low-mass}

\maketitle
\thispagestyle{headings}

\section
{Introduction}

Until recently, Gliese 229B (hereafter Gl 229B)
(Nakajima et al. 1995) had been the only
brown dwarf cool enough to show methane absorption
in its spectrum (Oppenheimer et al. 1995). It is still the only
cool brown dwarf companion to a nearby star.
In 1999, two large scale sky surveys, the Sloan Digital
Sky Survey (Strauss et al. 1999) and the 2MASS (Burgasser et al. 1999)
discovered six methane brown dwarfs in the field.
Detailed studies of methane brown dwarfs, or
T dwarfs (Kirkpatrick et al. 1999),  are now becoming
possible. Cuby et al. (1999) have found a very faint methane
brown dwarf, despite  observational difficulty.

The only T dwarf whose optical and
infrared spectra have been studied in detail
is Gl 229B (Geballe et al. 1996, Oppenheimer et al. 1998).
Optical and infrared spectra have been obtained
of SDSS 1624+00
(Strauss et al. 1999) and low-resolution infrared spectra
have been obtained of the four 2MASS T dwarfs
(Burgasser et al. 1999). 
The spectrum of Gl 229B is characterized by
broad methane and water vapor absorption bands (Oppenheimer 1995).
The details of methane and water vapor features were
studied by Geballe et al. (1996).

Two groups have reported the detection of CO in Gl 229B
(Noll et al. 1997, Oppenheimer et al. 1998).
The presence of CO is not consistent with the cool environment of
the T dwarf, and some mechanism that brings up CO to upper
atmosphere from a deep hot region must exist.
As for atomic lines,
Oppenheimer et al. (1998)
identified two Cs I lines in the far optical spectrum
of Gl 229B. One of the Cs I lines is also seen the spectrum
of SDSS 1624+00 by Strauss et al. (1999)  
who have found that two K I lines in the J band
are stronger in SDSS 1624+00 than Gl 229B. 

Weather has been searched 
for in a couple of warm brown dwarfs by
spectral variability (Tinney and Tolley 1999). 
A large number of warm brown dwarfs are seen to be
rapidly rotating (Mart\'{\i}n et al. 1997, Tinney and Reid 1998),
and rotation periods expected from rotation velocities
are 3$-$6 h, comparable to the rotation period of the Jupiter.
So it seems reasonable to assume that the rotation periods
of cool brown dwarfs are of the same order. 
Tinney and Tolley did not find any variability
of the L-type brown dwarf DENIS-PJ1228-1547, but found the variability
of the M-type brown dwarf LP 944-20.

In this paper, we report the results of
a search for spectral variability
of SDSS 1624+00 in the near infrared.
An integrated spectrum with high signal-to-noise ratio 
is used for detailed investigation of spectral features
in comparison with the previously obtained Gl 229B spectra
by Geballe et al. (1996) and Oppenheimer et al (1998).

\section{Observations and Data Reduction}

Observations were carried out on 1999 June 20 (UT) on the Subaru
Telescope,
using the grism mode of the Cooled Infrared Spectrograph and Camera
for OH Suppression Spectrograph (CISCO)
which employs a 1024$\times$1024 HgCdTe array.
The slit width of $0.\hspace{-2pt}''7$ corresponds to
a resolution of 0.005 $\mu$m in the wavelength range from 1.05 to
1.8 $\mu$m.   
The pixel scale of $0.\hspace{-2pt}''116$/pix gives six pixels 
per resolution.
The sky was almost clear, but the image size varied between 0.4 and
$1\hspace{-1pt}''$.  
We obtained 27 useful 100 second exposures
between 10:53 and 12:13 UT, which were once interrupted 
by a telescope tracking error.
The log of observing is given in table 1.


\begin{table}
\small
\begin{center}
Table~1.\hspace{4pt} Log of observations. \\
\end{center}
\vspace{6pt}
\begin{tabular}{ccc}
\hline\hline
Spectrum  &  Start time (1999 June 20 UT)  & Comment \\
\hline 
1 &	10:53:03 & \\
2 &	10:54:47 & \\
3 &	10:56:32 & \\
4 &	10:58:17 & \\
5 &	11:32:53 & Interruption before observing \\
6 &	11:34:37 & \\
7 &	11:36:22 & \\
8 &	11:38:07 & \\
9 &	11:39:52 & \\
10 &	11:42:18 & \\
11 &	11:44:03 & \\
12 &	11:45:48 & \\
13 &	11:47:32 & \\
14 &	11:49:17 & \\
15 &	11:51:02 & \\
16 &	11:53:22 & \\
17 &	11:55:07 & \\
18 &	11:56:52 & \\
19 &	11:58:37 & \\
20 &	12:00:22 & \\
21 &	12:02:07 & \\
22 &	12:04:27 & \\
23 &	12:06:12 & \\
24 &	12:07:58 & \\
25 &	12:09:43 & \\
26 &	12:11:28 & \\
27 &	12:13:12 & \\
\hline
\end{tabular}

\end{table}

Within the slit length of $119\hspace{0pt}''$, 
a nearby bright star $56.\hspace{-2pt}''7$
away
was observed simultaneously to the target brown dwarf.
For sky cancellation,
both the object and the nearby star were shifted
along the slit. 
This nearby star was used to monitor the possible variation of
transmission. Apart from the nearby star, the A0 star SAO121542
(V = 7.7) and
the F2 star SAO121611 (V = 9.0), were observed for reference after
the target observations. 

IRAF was used for data reduction.
Each frame was sky subtracted and then
divided by the average dome flat frame.
The flattened frame was 
distortion corrected and wavelength calibrated, using
OH airglow lines. Then the spectra of SDSS 1624+00 and
the nearby star were extracted. 

The spectral type of the nearby bright star is unknown.
By the comparison of the continuum slopes of the nearby
star,  A0 star, and F2 star, the effective temperature 
of the nearby star was estimated to be 4800K. This
estimate is consistent with the presence 
of weak Pa$\beta$ absorption in the nearby star spectrum.

No photometry for absolute flux calibration was obtained.
The relative flux of each object spectrum was calibrated
by dividing by the nearby star spectrum in the same
frame and multiplying by a 4800K blackbody spectrum. 
In figure 1, the ratio of the first nearby star spectrum
 to the last is plotted. From the figure, it is apparent
that transmission variation due to telluric water vapor
was significant over the time scale of 80 minutes, and that
the calibration by the simultaneously observed nearby star
was essential.


\section{Discussion}

\subsection{Variability}

In order to discuss variability, one must know the level of
noise in each spectrum obtained in 100 seconds. 
The major source of noise is shot noise in
OH airglow lines. To see the influence of the OH lines, a sky spectrum
was extracted using the same aperture for object
and its square root was calculated. 
The square root of the sky spectrum is shown in figure 2
along with a typical raw spectrum of SDSS 1624+00 which is
not calibrated for wavelength and flux.


We have inspected calibrated spectra for 100 second exposures
and found some possible variations,
but they are not definitive  in view of the
limited signal-to-noise ratio. 
To see longer term variability and to improve the signal-to-noise
ratio, four or five spectra were coadded
to produce one spectrum corresponding to several minutes of observing.
The combination of the spectra is shown in table 2.

\begin{table}
\small
\begin{center}
Table~2.\hspace{4pt} Combined spectra. \\
\end{center}
\vspace{6pt}
\begin{tabular}{cc}
\hline\hline
Combined spectrum & Original Spectra \\
\hline
A & 1, 2, 3, 4 \\
B & 5, 6, 7, 8, 9 \\
C & 8, 9, 10, 11, 12 \\
D & 11, 12, 13, 14, 15 \\
E & 14, 15, 16, 17, 18 \\
F & 17, 18, 19, 20, 21 \\
G & 20, 21, 22, 23, 24 \\
H & 23, 24, 25, 26, 27 \\
I & 1 -- 27 \\
\hline
\end{tabular}

\end{table}

Time series of the spectra denoted by  A through H
and the integrated spectrum  denoted by I are shown in figure 3. 
The observations of A were more than 30 minutes
before B due to interruption by the telescope tracking problem.
Two types of small variation are seen in the time series.
One is seen between A and the rest of the spectra, and the other
is found throughout the time span.
The major difference between A and the rest is
the absorption features at 1.066 and 1.081 $\mu$m.
These features are seen in three of the four original 100 second
spectra. 
The left shoulder of the H band pseudo-continuum appears to vary
continuously. The sawtooth pattern which extends from 1.53
to 1.58 $\mu$m in the spectra E and F is not clear in A and
H. The pattern gradually grows to the left from A to F
and becomes less obvious from F to H.
To emphasize this variation, the average spectrum of A and H
and that of E and F are compared in figure 4 in which vertical
dotted lines indicate the locations of the three valleys of
the sawteeth. 
%
%
It is likely that at least the latter type of variation is due to water vapor.
To confirm the presence of this variability, we plan
observations with a longer time baseline.
The possible variation at 1.668 $\mu$m coincides with the location of
a strong OH line and its reality is doubtful.

\subsection{Integrated Spectrum}

Here we discuss the details of the integrated spectrum 
in comparison with the previously obtained spectra of Gl 229B
by Geballe et al. (Geballe et al. 1996, Leggett et al. 1999)  
and Oppenheimer et al. (1998)
and with the spectrum of SDSS 1624+00 by
Strauss et al. (1999).
Geballe et al. and Strauss et al. obtained their
spectra with CGS4 on UKIRT and Oppenheimer et al. 
with NIRC on Keck.

One major difference between Gl 229B and SDSS 1624+00 is that
the former is brighter by 1.2 mag. 
Strauss et al. emphasize
the similarity in the spectra of the two methane brown dwarfs and
assume the same effective temperature and luminosity to
drive the distance to SDSS 1624+00.
However we have found from an analysis of model spectra described
later that
the overall near-infrared spectrum does not change significantly
in the temperature range between 900K and 1100K apart from the absolute
flux level which is unknown for SDSS 1624+00.
So individual spectral features which are sensitive to temperature
and dust need to be utilized.

Our integrated spectrum I of SDSS 1624+00 and
the Gl 229B spectrum by Geballe et al. are plotted in figure 5
in log $f_\nu$. Strauss et al. find that the significant 
difference is the excess of flux around 1.7 $\mu$m in SDSS 1624+00
and the stronger lines of K I 1.2432 and 1.2522 $\mu$m.
Our findings are basically the same, the deeper methane feature
including the region around 1.7 $\mu$m in Gl 229B and the stronger
K I lines in SDSS 1624+00. 
In addition from the Gl 229B spectrum of Oppenheimer et
al., we find that the water vapor absorption bands 
at 1.15 $\mu$m and 1.4 $\mu$m are deeper in Gl 229B.
The signal-to-noise ratio of the water vapor absorption
bands of the Gl 229B spectrum of Geballe et al. is limited, but
that of Oppenheimer et al. is valid to the edges of bottoms
of the absorption bands. 
From the analysis of
the model spectra, we have found that these features are sensitive
to temperature and the presence of dust.


In figure 6, six model spectra are plotted for
dusty and dust free models for temperatures, 900, 1000K and 1100K.
Here, the dust free model describes the situation that dust has formed, but
has been segregated from the gas throughout the photosphere,
and hence plays little role as a source of opacity. On the
other hand, in the dusty model, dust is active as a source of
opacity in the warm region deep in the photosphere 
(Tsuji et al. 1999).
The models show following tendencies for
the above mentioned spectral features. The methane absorption
is deeper at the lower temperature and the K I lines are  stronger at the
higher temperature. 
Therefore the methane absorption indicates that Gl 229B is
cooler, and the K I lines suggest that SDSS 1624+00 is warmer.
From above information alone, one may judge that SDSS 1624+00
is warmer and the atmosphere is dust free.
However, the situation is more complicated. 
We have detected
a finite flux at the bottom of water vapor absorption
between 1.34 and 1.42 $\mu$m in SDSS 1624+00, 
and the water vapor absorption
is deeper in Gl 229B. According to the models, the water vapor
absorption is deeper at the lower temperature and shallower 
in the dustier object. 
The dust free model tends
to produce excessively deep water vapor absorption in SDSS 1624+00
and the shallowness
of the trough is more consistent with the dusty model.
From the spectral features, we conclude that
SDSS 1624+00 is warmer and dustier than Gl 229B. 
Parallax measurements of SDSS 1624+00 is extremely important 
in determining the effective temperature.

K I 1.2432 and 1.2522 $\mu$m lines 
corresponding to  the transition array, 4{\it p}--5{\it s}
(multiplet $^2$P$^0$--$^2$S) (Wiese et al. 1966), are
seen in both spectra.  
So it is natural to expect that
the lines for the transition array, 4{\it p}--3{\it d} should
be seen, because of the similar excitation and high transition
probabilities. In the spectrum I, we have identified two lines
at 1.1690 and 1.1773 $\mu$m corresponding to the transition array,
4{\it p}--3{\it d}. In the spectrum by Strauss et al. the 1.1773
$\mu$m line is clearly detected and the 1.1690 $\mu$m is weakly
detected though the signal-to-noise ratio is lower.
However these lines are not apparent
in the Gl 229B spectrum by Geballe et al. (1996).
These lines have been detected in late 
M dwarfs by Kirkpatrick et al. (1993).

\section{Summary}

We have searched for spectral variability of SDSS 1624+00
with a 500 second time resolution over more than an hour.
The suspected variation was found in water vapor absorption. 
Observations with a longer time baseline
will be needed to confirm this variability.
The integrated spectrum was compared with the previously
obtained Gl 229B spectra
and the model spectra.
Methane absorption, the KI lines, and water vapor absorption 
indicate that SDSS 1624+00 is warmer and
dustier than Gl 229B.
Absolute flux calibration
by parallax measurements of SDSS 1624+00 is needed to clarify
the effective temperature problem.

We thank T. Geballe, M. Strauss, and B. Oppenheimer for providing
their spectra in electronic form. We thank B. Oppenheimer,
M. Iye and the referee, D. Kirkpatrick for 
useful comments
on the manuscript.
We thank the staff of the Subaru
Observatory for the support of observing. 
This research is supported by Grants-in-Aids for
Scientific Research of the Japanese Ministry of
Education, Culture, Sports, and Science
(No. 10640239 to TN and No. 11640227 to TT)
and by
the Sumitomo Foundation (TN).

\vskip 1cm

\section*{References}

\re
Burgasser A. J., Kirkpatrick J. D., Brown M. E., Reid I. N.,
Gizis J. E., Dahn, C. C., Monet D. G., Beichman C. A. et al.
1999, ApJ 522, L65

\re
Cuby J. G., Saracco P., Moorwood, A. F. M., D'Odorico S,
Lidman C, Comeron F., Spyromilio J. 1999, A\&A 347, L41 

\re
Geballe T. R., Kulkarni S. R., Woodward C. E., Sloan G. C. 1996,
ApJ 467, L101

\re
Kirkpatrick J. D., Kelley D. M., Rieke G. H., Liebert J.,
Allard F., Wehrse R. 1993, ApJ 402, 643

\re
Kirkpatrick J. D., Reid, I. N., Liebert, J., Cutri, R. M.,
Nelson, B., Beichman, C. A., Dahn, C. C., Monet, D. G. et al. 
1999, ApJ 519, 802

\re
Leggett S. K., Toomey D. W., Geballe T. R., Brown R. H. 1999,
ApJ 517, L139

\re
Mart\'{\i}n E. L., Basri. G., Delfosse X., Forveille T. 1997,
A\&A 327, L29 

\re
Nakajima T., Oppenheimer B. R., Kulkarni S. R., Golimowski D. A.,
Matthews, K., Durrance S. T. 1995, Nature 378, 463 

\re
Noll K. S., Geballe T. R., Marley M. S. 1997, ApJ 489, L87

\re
Oppenheimer B. R., Kulkarni S. R., Matthews K., Nakajima T. 1995,
Science 270, 1478

\re
Oppenheimer B. R., Kulkarni S. R., Matthews K., van Kerkwijk M. H.
1998, ApJ 502, 932

\re
Strauss M. A., Fan X., Gunn J. E., Leggett S. K., 
Geballe T. R., Pier, J. R., Lupton R. H., Knapp G. H. et al.
1999, ApJ 522, L61

\re
Tinney C. G., Reid I. N. 1998, MNRAS 301, 1031 

\re
Tinney C. G., Tolley A. J. 1999, MNRAS 304, 119 

\re
Tokunaga  A. T., Kobayashi N. 1999, AJ 117, 1010

\re
Tsuji T., Ohnaka K., Aoki W. 1999,
ApJ 502, L119

\re
Wiese, W. L., Smith, M. W., Glennon, B. M., 1966,
Atomic Transition Probabilities, Vol. 1. (Washington, D. C.; GPO)

\clearpage
\centerline{Figure Captions}
\bigskip

\vskip 1cm
\noindent
Figure 1.
{Transmission variation. The ratio of the first nearby star spectrum
to the last
is plotted as a function of wavelength. From the figure,
it is apparent that transmission variation due to telluric
water vapor was significant over the time scale of 80 minutes,
and that the calibration by the simultaneously observed nearby
star was essential.}

\vskip 1cm
\noindent
Figure 2.
{Signal to noise ratio in a raw spectrum. A typical raw spectrum
of SDSS 1624+00 in electron count (solid line)
is plotted along with the square root of
the sky spectrum extracted with the same aperture as the object
(dotted line).
The ratio of the two gives the signal-to-noise ratio obtained
with a 100 second exposure.}

\vskip 1cm
\noindent
Figure 3.
{Times series of combined spectra. Each spectrum
corresponds to several minutes of observing. 
The spectra are denoted by A through H from the bottom to the 
eighth in the order 
of observing time. The top spectrum is
the  integrated spectrum  denoted by I. }

\vskip 1cm
\noindent
Figure 4.
{Spectral variability. The average spectrum of A and H in figure 3
is compared with that of E and F. The valleys of the sawtooth pattern
seen in E+F is not obvious in A+H. This variation is likely due to
water vapor.}

\vskip 1cm
\noindent
Figure 5.
{Comparison of spectra of SDSS 1624+00 and Gl 229B.
Overall Gl 229B is brighter by 1.2 mag. The Gl 229B
spectrum by Geballe et al. was obtained with a higher resolution.
The methane absorption longward of 1.62 $\mu$m is shallower
in SDSS 1624+00 and KI lines at 1.2432 and 1.2522 $\mu$m are
stronger. Finite emission is seen at the bottom of
water vapor absorption between 1.34 and 1.42 $\mu$m
in the SDSS 1624+00 spectrum. 
} 

\vskip 1cm
\noindent
Figure 6.
{Model spectra. D and F denote dusty and dust-free models respectively
and the number indicates the effective temperature.
The model spectra show some tendencies of spectral features.
In the cooler model, methane and water vapor absorption
are deeper and the KI lines are weaker. In the dusty
model, water vapor absorption is shallower.
}

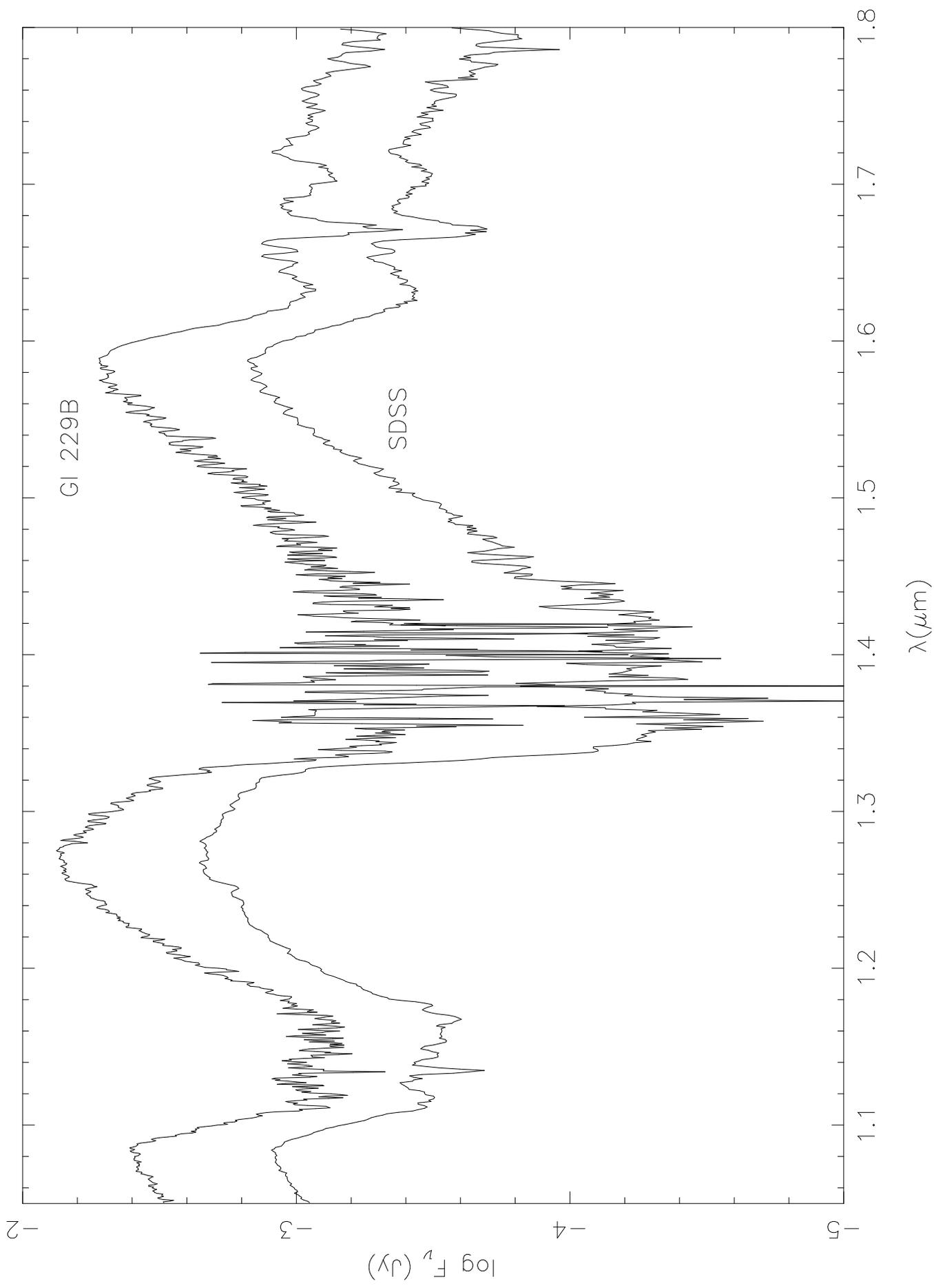
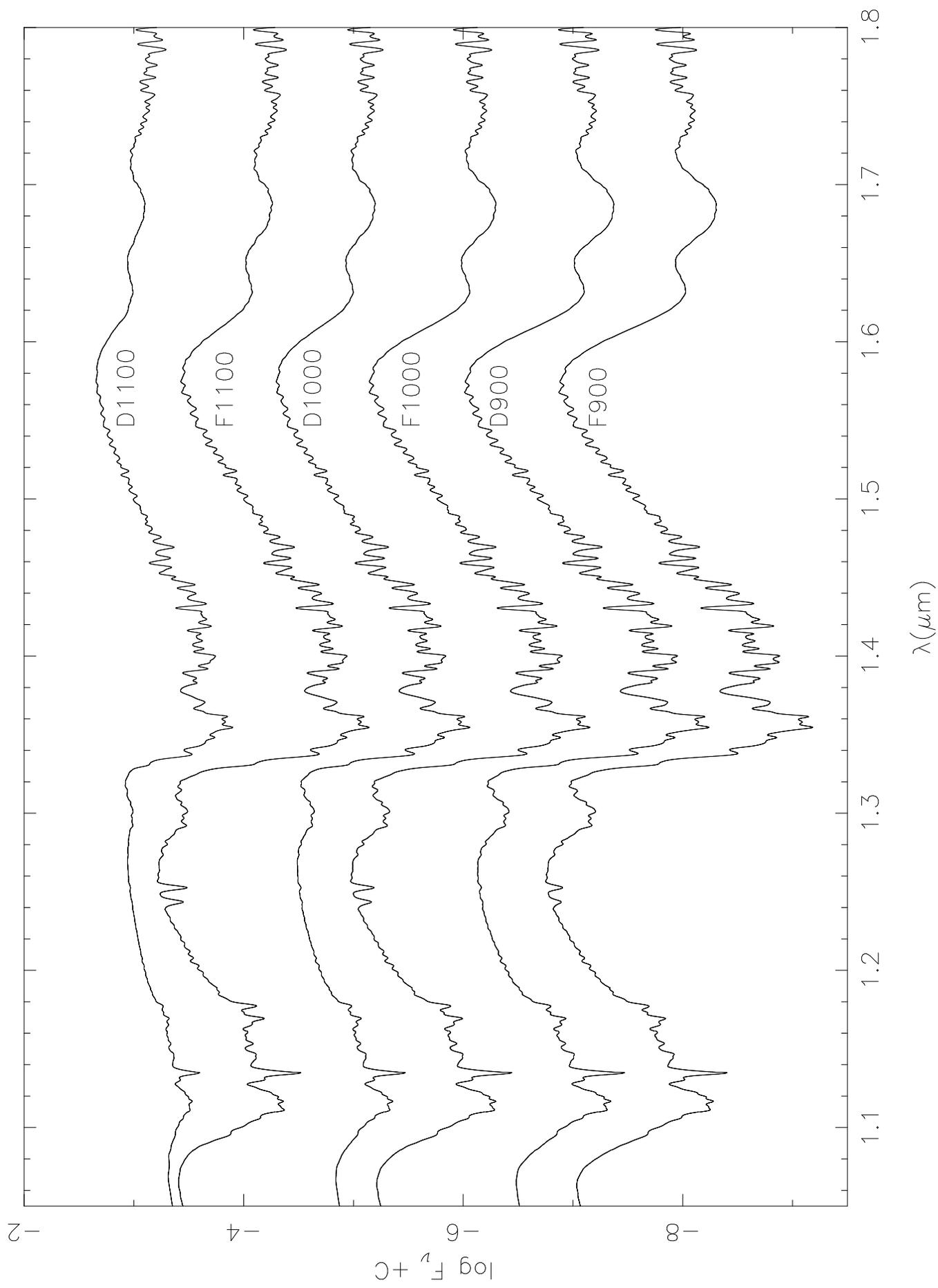
\end{document}